\documentstyle[12pt]{article}

\begin{document}
\begin{center}
\begin{Large}

{\bf Plasmon dispersion in quasi-one and one-dimensional systems
with non-magnetic impurities}
\end{Large}
\vspace{1.5cm}

\begin{large}
{\bf I. Grosu and L.Tugulan}
\end{large}
\vspace{0.5cm}

{\bf Department of Theoretical Physics, University of Cluj, 3400 Cluj,
Romania}
\end{center}
\vspace{1cm}

We calculate the plasmon dispersion in quasi-one-dimensional quantum
wires, in the presence of non-magnetic impurities, taking into
consideration the memory function formalism and the role of the
forward scattering. The plasma frequency is reduced by the presence
of impurities. We also calculate, analytically, the plasmon
dispersion in the Born approximation, for the scattering of the
electrons by the non-magnetic impurities. We compare our result with
the numerical results of Sarma and Hwang. \vspace{1cm}

{\bf Keywords:} Quasi-one-dimensional systems, Disorder, Forward
scattering, Plasmon modes, Memory function formalism, Born
approximation
\newpage

\begin{large}
{\bf 1. Introduction}
\end{large}
\vspace{0.5cm}

The problem of the existence of one-dimensional plasmon modes was
discussed by Ritsko et.al.[1] after the measurements on
one-dimensional organic metals by high-energy inelastic electron
scattering in thin crystalline films. Short after, Williams and
Bloch [2] developed a random-phase-approximation (RPA) model, in
order to calculate the dielectric response of a
quasi-one-dimensional metal, with the result in a good agreement
with the experimental data. They also predict an acoustic-plasmon
branch. The progress in semiconductor fabrication allowed the
experimentalists for the realization of systems where electrons are
confined in one dimension [3]. The measurements were performed on
GaAs quantum wires, using the inelastic resonant light scattering
spectroscopy that allow for the investigation of the collective and
single-particle excitations of the electron gas. The measured
wave-vector dependence of the plasmon dispersion was in qualitative
agreement with the random-phase-approximation calculations. Li
et.al.[4], showed that the RPA provides an exact description of the
plasmon dispersion in one-dimensional wires, by establishing an
equivalence between RPA and Tomonaga-Luttinger liquid model [5,6].
The dynamical response of an one-dimensional electron system in a
quantum-wire was theoretically and numerically analyzed, within RPA,
by Sarma and Hwang [7]. They also included the effect of disorder,
in the Born approximation, using a numerical procedure in order to
find the modified plasma frequency. The effect of the
electron-electron correlations on the scattering processes in
quantum wires was discussed by Thakur and Neilson [8]. The
scattering rate increases with the increasing of the electron
correlations, and the plasmon dispersion depends on the electron
correlations and on the level of disorder. Other measurements, on
quasi-one-dimensional organic conductors [9], established the
frequency dependence of the conductivity, that can be explained
using the role of the forward scattering by impurities [10]. In this
paper, we will analyze the plasmon dispersion in
quasi-one-dimensional systems in the presence of disorder, beyond
the Born approximation, using the memory function formalism. We also
calculate, analytically, the plasma frequency in the Born
approximation. The similarities and the differences between the
models are discussed. \vspace{0.5cm}

\begin{large}
{\bf 2. Model}
\end{large}
\vspace{0.5cm}

In order to calculate the quasi-one-dimensional plasma frequency, we
use a model that consists in a quasi-one-dimensional conductor with
disorder that can be modelled by the Gaussian potential $V(x)$, with
the average $\overline{V}(x)=0$, and the correlator given by:
$$\overline{V(x)\;V(x')}=\gamma_{0}\;C(x-x')\eqno{(1)}$$
where $\gamma_{0}$ is a measure of the strength of the disorder, and $C(x)$ is
defined in Ref.[10]. The plasmon dispersion will be analyzed investigating the
zeros of the complex dielectric function:
$$\epsilon(q,\omega)=1-V(q)\;\chi(q,\omega)\eqno{(2)}$$
Here $V(q)$ is the Fourier transform of the Coulomb interaction, and
$\chi(q,\omega)$ is the density-density response function. According to Ref.[11],
$\chi(q,\omega)$ can be expressed, in the long-wavelength limit, as:
$$\chi(q,\omega)\simeq -\frac{iq^{2}}{\omega e^{2}}\;\sigma(\omega)\eqno{(3)}$$
where $\sigma(\omega)$ is the frequency dependent conductivity. In
our model we take into consideration the effect of forward
scattering by impurities in an interacting electron system. Using
the memory function formalism, according to Ref.[12], the frequency
dependent conductivity is given by:
$$\sigma(\omega)=\frac{i D_{c}}{\omega+M(\omega)}\eqno{(4)}$$
Here $D_{c}$ is the charge stiffness, and $M(\omega)$ is the memory function
(which includes the vertex corrections to all orders in the perturbation
theory. $M(\omega)$ is expressed in powers of the impurity potential). The
memory function will be calculated using the Tomonaga-Luttinger model with
interaction parameters $g_{2}=g_{4}=\pi v_{F} F$, where $v_{F}$ is the Fermi
velocity, and $F$ is an interaction parameter, and with:
$$\Pi(q,\omega)=Z_{q}\;\frac{2\omega_{q}}{\omega_{q}^{2}-\omega^{2}}\eqno{(5)}$$
where:
$$Z_{q}=\frac{|q|}{\pi\sqrt{1+F}}\eqno{(6)}$$
and:
$$\omega_{q}=\overline{v}_{F}|q|\eqno{(7)}$$
(here $\overline{v}_{F}=v_{F}\sqrt{1+F}$). According to [10,12], the complex
memory function is given by:
$$M(\omega)\simeq\frac{2\gamma_{0}}{\pi^{2}nm^{*}\xi\overline{v}_{F}^{3}
\sqrt{1+F}}\;\omega+i\;\frac{\gamma_{0}}{\pi n m^{*}\overline{v}_{F}^{4}
\sqrt{1+F}}\;\omega^{2}\equiv b\;\omega+i\;a\;\omega^{2}\eqno{(8)}$$
with $m^{*}$ the band mass, and $n$ the density of the one-dimensional
electron gas. In Eq.(8), $p_{m}\sim\xi^{-1}$ is the maximal possible
momentum transfer between two electrons due to the scattering by the
impurity potential. When $\xi^{-1}\ll 2p_{F}$, the impurities do not give
rise to backward scattering, and the physics of the scattering is dominated
by the forward scattering. In this case, one can approximate the Fourier
transform of the factor $C(x)$, by the following form:
$$C(q)=\theta(1-|q|\xi)\eqno{(9)}$$
where $\theta(x)$ is the step function. Using Eqs.(8) and (4), the
frequency-dependent conductivity will be:
$$\sigma(\omega)=Re\;\sigma(\omega)+i\;Im\;\sigma(\omega)\eqno{(10)}$$
with:
$$Re\;\sigma(\omega)=\frac{\overline{D}_{c}\tau}
{1+(\omega\tau)^{2}}\eqno{(11)}$$
and:
$$Im\;\sigma(\omega)=\frac{\overline{D}_{c}}{\omega[1+(\omega\tau)^{2}]}
\eqno{(12)}$$
Here:
$$\overline{D}_{c}=\frac{D}{1+b}\eqno{(13)}$$
is the renormalized charge stiffness, and the time $\tau$ is given by:
$$\tau=\frac{a}{1+b}\eqno{(14)}$$
Due to the restrictions of the model, Eqs.(11) and (12) are valid only for
$\omega\tau<1$. With these results, the complex dielectric function
becomes:
$$\epsilon(q,\omega)=1-\frac{V(q)q^{2}\overline{D}_{c}}
{e^{2}\omega^{2}(1+\omega^{2}\tau^{2})}+i\;\frac{V(q)q^{2}\overline{D}_{c}
\tau}{e^{2}\omega(1+\omega^{2}\tau^{2})}\eqno{(15)}$$
The plasma frequency is determined by the zeros of the dielectric function:
$$\epsilon(q,\omega_{p})=0\eqno{(16)}$$
With the condition $\omega_{p}\tau<1$, Eq.(16) reduces to:
$$e^{2}\omega_{p}^{2}+iV(q)q^{2}\overline{D}_{c}\tau\omega_{p}-
V(q)q^{2}\overline{D}_{c}\simeq 0\eqno{(17)}$$
The stable solution is given by:
$$\omega_{p}=\frac{V(q)q^{2}\overline{D}_{c}\tau}{2e^{2}}
\left[\sqrt{\frac{4e^{2}}{V(q)q^{2}\overline{D}_{c}\tau^{2}}-1}-i\right]
\eqno{(18)}$$
In the absence of disorder, and in the long-wavelength limit, it was shown [4]
that the plasma frequency can be calculated as:
$$\omega_{p}\sim q\;|\ln(qa)|^{1/2}\eqno{(19)}$$
where $V(q)\sim|\ln(qa)|$ is the Fourier transform of the Coulomb
interaction. Here $a$ is a typical confinement width (introduced in
order to avoid the logarithmic divergence, typical to Coulomb
interaction in a strictly one dimensional system). $Re\;\omega_{p}$
increases from zero with the increasing of the wavevector $q$. There
is no overdamping for small wavevectors (In two dimensional case
[13], and for one dimensional systems in the Born approximation [7],
there is an overdamping for small wavevectors). For small
wavevectors $q$, $|Im\;\omega_{p}|$ is a slightly increasing
function. The presence of non-magnetic impurities reduces the
electron plasma frequency. A similar behavior was obtained in
Ref.[7], in the Born approximation. When the value of the wavevector
$q$ increases, $Re\;\omega_{p}$ becomes of the order of
$|Im\;\omega_{p}|$ and the plasmon mode becomes strongly damped.
\vspace{0.5cm}

\begin{large}
{\bf 3. Analytical results in the Born approximation}
\end{large}
\vspace{0.5cm}

In this section, following Ref.[7], we will give an analytical
result for the plasmon frequency in the presence of randomly
distributed impurities, and in the Born approximation. In this case
the Green's function becomes [14]:
$$G(q,\omega)=\frac{1}{\omega-\epsilon(q)+\frac{i}{2\tau}sign(\omega)}\eqno{(20)}$$
where $\tau$ is the elastic scattering time. In this case, the
polarizability $\Pi(q,\omega)$ is given by:
$$\Pi(q,\omega)=N_{0}\frac{Dq^{2}/\tau}{(\omega+i/\tau)\left[\omega-\frac{i}{\tau}
\frac{Dq^{2}/\tau}{(\omega+i/\tau)^{2}}\right]}\eqno{(21)}$$ Here
$N_{0}$ is the density of states per spin at the Fermi level, and
$D=v_{F}^{2}\tau$ is the diffusion coefficient. The Coulomb
interaction potential, for $qa\ll 1$, is:
$$V(q)=\frac{2e^{2}}{\epsilon_{0}}\;K(|qa|)\simeq
-\frac{2e^{2}}{\epsilon_{0}}\;\ln|qa|\eqno{(22)}$$ The plasmon
frequency is given by the zeros of the dielectric function. In this
case, and in the approximation $(\omega_{p}\tau)^{2}\ll 1$, the
equation for $\omega_{p}$ becomes:
$$1-\frac{f(q)\left(1-i\omega_{p}\tau\right)}{Dq^{2}-i\omega_{p}-2\omega_{p}^{2}\tau}\simeq
0\eqno{(23)}$$ where:
$$f(q)=\frac{2Dq^{2}e^{2}N_{0}\ln|qa|}{\epsilon_{0}}\eqno{(24)}$$
In the long-wavelength limit, the solution of Eq.(23) is:
$$\omega_{p}\simeq\frac{1}{4\tau}\sqrt{8 Dq^{2}\tau-6\alpha
Dq^{2}\tau\ln|qa|-1}-i\;\frac{1-\alpha Dq^{2}\tau
\ln|qa|}{4\tau}\eqno{(25)}$$ with
$$\alpha=\frac{2e^{2}N_{0}}{\epsilon_{0}}\eqno{(26)}$$
For wavevectors $q$ between 0 and $q_{c}$ the plasmon modes are
overdamped. Here $q_{c}$ is a critical wavevector, given by the
solution of the following equation:
$$8 q^{2}-6\alpha q^{2} \ln|qa|-\frac{1}{D\tau}=0\eqno{(27)}$$
The critical wavevector is:
$$q_{c}=\frac{1}{\sqrt{D\tau}}\sqrt{-\frac{1}{3\alpha\;W(-1,
-b)}}\eqno{(28)}$$ Here $W(-1, -b)$ is the $-1$ branch of the
Lambert function, and $b$ is given by:
$$b=\frac{a^{2}}{3\alpha
D\tau}\;\exp\left(-\frac{8}{3\alpha}\right)\eqno{(29)}$$ From
Eqs.(25) and (28) we will recover, analytically, the conclusions of
Sarma and Hwang [7], obtained numerically in the Born approximation.
The motivation of such an estimation was pointed in Ref.[7] for the
case of high mobility GaAs quantum wire systems, where the
localization lengths are very large. \vspace{0.5cm}

\begin{large}
{\bf 4. Conclusions}
\end{large}
\vspace{0.5cm}

We calculated the electron plasma frequency for a
quasi-one-dimensional system in the presence of disorder introduced
by non-magnetic impurities, taking into consideration of the role of
the forward scattering and using the memory function formalism. In
this model the presence of non-magnetic impurities reduces the
plasma frequency. This result is typical for 1D and 2D systems where
the scattering by the impurities strongly affect the plasmon
dispersion [7,13]. We also calculate the plasma frequency for an
one-dimensional system, in the presence of disorder. We use the Born
approximation that describes the scattering of the electrons by the
non-magnetic impurities [14]. The analytical result is in agreement
with the numerical result of Sarma and Hwang. It was shown that, in
the Born approximation, the plasmon modes are overdamped in the long
wavelength limit. On the other hand, in our model from Sec.2, there
is no overdamping in the $q\rightarrow 0$ limit. This is a new
result, and this behavior is due to the consideration of the forward
scattering in the memory function formalism. However, in such
systems, backward scattering are also present, and in pure
one-dimensional systems can lead to localization and to a vanishing
conductivity for small frequencies. In quasi-one-dimensional systems
there can be an interchain coupling that can stabilize a
quasimetallic phase. This gives a nonvanishing small frequencies
conductivity that could be fitted by a formula given in Ref.[9], and
theoretically explained in the model of forward scattering. However,
it is difficult to establish an experimental support for the model,
because of the experimental difficulties and the narrow interval (in
disorder) of the validity of the model. \vspace{1cm}

\begin{large}
{\bf References}
\end{large}
\vspace{1cm}

[1] J.J.Ritsko, D.J.Sandman, A.J.Epstein, P.C.Gibbons, S.E.Schnatterly,

J.Fields, Phys.Rev.Lett.34, 1330, (1975)

[2] P.F.Williams, A.N.Bloch, Phys.Rev.Lett.36, 64, (1976)

[3] A.R.Goni, A.Pinczuk, J.S.Weiner, J.M.Calleja, B.S.Dennis, L.N.Pfeiffer,

K.W.West, Phys.Rev.Lett.67, 3298, (1991)

[4] Q.P.Li, S.Das Sarma, R.Joynt, Phys.Rev.B 45, 13713, (1992)

[5] S.Tomonaga, Prog.Theor.Phys.5, 544, (1950)

[6] J.M.Luttinger, J.Math.Phys.4, 1154, (1963)

[7] S.Das Sarma, E.H.Hwang, Phys.Rev.B 54, 1936, (1996)

[8] J.S.Thakur, D.Neilson, Phys.Rev.B 56, 4679, (1997)

[9] A.Schwartz, M.Dressel, G.Gr\"{u}ner, V.Vescoli, L.Degiorgi, T.Giamarchi,

Phys.Rev.B 58, 1261, (1998)

[10] P.Kopietz, G.E.Castilla, Phys.Rev.B 59, 9961, (1999)

[11] D.Pines, P.Nozieres, "The theory of quantum liquids", vol.1,
Benjamin,

New York, (1966)

[12] W.G\"{o}tze, P.W\"{o}lfle, Phys.Rev.B 6, 1226, (1972)

[13] G.F.Giuliani, J.J,Quinn, Phys.Rev.B 29, 2321, (1984)

[14] G.Rickayzen, "Green's functions and condensed matter",

Academic Press, New York, (1980)

\end{document}